# Terahertz radiation by subpicosecond spin-polarized photocurrent originating from Dirac electrons in a Rashba-type polar semiconductor

Yuto Kinoshita[1], Noriaki Kida[1,*], Tatsuya Miyamoto[1], Manabu Kanou[2], Takao Sasagawa[2], and Hiroshi Okamoto[1,3]

[1]*Department of Advanced Materials Science, The University of Tokyo, 5-1-5 Kashiwa-no-ha, Chiba 277-8561, Japan*

[2]*Materials and Structures Laboratory, Tokyo Institute of Technology, Kanagawa 226-8503, Japan*

[3]*AIST-U. Tokyo Advanced Operando-Measurement Technology Open Innovation Laboratory, National Institute of Advanced Industrial Science and Technology, Chiba 277-8568, Japan*

* kida@k.u-tokyo.ac.jp

The spin-splitting energy bands induced by the relativistic spin-orbit interaction in solids provide a new opportunity to manipulate the spin-polarized electrons on the sub-picosecond time scale. Here, we report one such example in a bulk Rashba-type polar semiconductor BiTeBr. Strong terahertz electromagnetic waves are emitted after the resonant excitation of the interband transition between the Rashba-type spin-splitting energy bands with a femtosecond laser pulse circularly polarized. The phase of the emitted terahertz waves is reversed by switching the circular polarization. This suggests that the observed terahertz radiation originates from the subpicosecond spin-polarized photocurrents, which are generated by the asymmetric depopulation of the Dirac state. Our result provides a new way for the current-induced terahertz radiation and its phase control by the circular polarization of incident light without external electric fields.

A helical Dirac cone realized in the surface state of topological insulators provides a new platform to study the spin-dependent light-matter interaction and also to fabricate spin-dependent optical devices [1]. For example, it is possible to selectively excite momentum-dependent spin-polarized Dirac electrons via spin-flipping process by an irradiation of a circularly polarized light so as to satisfy the optical selection-rule based upon the conservation of energy and angular-momentum [2]. Such a photoexcitation results in the asymmetric depopulation of the Dirac cone in the momentum space and thus the spin-polarized electrical currents are generated. This is known as the circular photogalvanic effect, which was originally found in a noncentrosymmetric gyrotoropic medium [3] and is now observed in a two-dimensional quantum well [4-6] and a voltage-biased nanowire [7].

The photocurrent $J_{\text{CPGE}}$ originating from the circular photogalvanic effect can be expressed as

$$J_{\text{CPGE}} \propto \gamma \hat{e} \sigma_{\pm} E^2. \qquad (1)$$

Here, $\gamma$ is a second-rank tensor introduced by the Rashba Dresselhaus spin-orbit interaction, $\hat{e}$ is a unit vector parallel to both the plane of incidence and the sample surface, $\sigma_{\pm} (= \pm 1)$ is the helicity of a circularly polarized light, and $E$ is the electric field of light [2]. In a topological insulator of $Bi_2Se_3$, $J_{\text{CPGE}}$ generated by an irradiation of a circularly polarized light was measured using a femtosecond laser pulse [8,9] and a continuous wave (cw) laser light [8] as excitation sources. Transient photocurrents can also be detected by the radiation measurement, since time-dependent currents generate electromagnetic waves via the electric-dipole radiation mechanism, which is expressed by

$$E_{\text{THz}} \propto \frac{\partial}{\partial t} J. \qquad (2)$$

Here, $E_{\text{THz}}$ is the radiated electric field and $J$ is the time-dependent current. Frequencies of such radiated electromagnetic waves are usually located in the terahertz region, so that this process is used as a typical method of terahertz radiation in semiconductors, e.g., in narrow-gap semiconductors such as InAs and InSb [10], and in voltage-biased photoswitching devices made on low-temperature-grown GaAs. In various topological insulators such as $Bi_2Se_3$ [9, 11-14], Cu-doped $Bi_2Se_3$ [11, 14], Ca-doped $Bi_2Se_3$ [15], and $Bi_{1.5}Sb_{0.5}Te_{1.7}Se_{1.3}$ [16] and graphene [17-19], light-induced terahertz radiations have also been observed. In those materials, however, a spin-polarized photocurrent originating from the asymmetric depopulation of the Dirac cone or equivalently the circular photogalvanic effect is excluded as an origin of the terahertz radiation. If a terahertz radiation is caused by the circular photogalvanic effect, the phase of the terahertz electromagnetic field can be controlled by the choice of the helicity of circular polarization without external electric fields. To achieve such an all optical control of terahertz radiation, we focus on a Rashba-type polar compound, BiTeBr, which is nonmagnetic (*n*-type) semiconductor.

Figure 1(a) shows the crystal structure of BiTeBr. The crystal system is trigonal and the space group is *P*3*m*1 (point group is $C_{3v}$). Three kinds of constituent ions, $Bi^{3+}$, $Te^{2-}$, and $Br^-$, respectively form plane units, which are stacked along the [001]-direction. Such a layered stacking of different planes breaks the space-inversion symmetry. In BiTeBr, it was reported that large spin-splitting bands with the splitting energy $E_{\text{split}}$ of ~0.1 eV appear by the Rashba-type spin-orbit interaction of Bi ions [20]. The shift of the conduction band minima $k_{\text{min}}$ from the A point in the Brillouin zone is estimated to be ~±0.043 Å$^{-1}$ [20]. The splitting bands with helical spins in BiTeBr are schematically shown in Fig. 1(b) and the helical spin textures at the Fermi-surface in the momentum space are illustrated in Fig. 1(c).

In the present study, we investigated the terahertz radiation from BiTeBr by the irradiation of a femtosecond laser pulse. We found that the phase of the radiated terahertz wave is inverted when the helicity of a circularly polarized light is switched, suggesting that the terahertz radiation originates from the asymmetric depopulation of the Dirac state. From the results of the terahertz electric-field waveforms as functions of polarization, fluence, and photon energy of the incident femtosecond laser pulse, we discuss the terahertz-radiation mechanism in detail.

[001]-oriented single crystals of BiTeBr were grown by the Bridgman method, details of which were reported elsewhere [21]. The flesh (001)-surface was obtained by cleaving an as-grown single crystal with the scotch tape. In the terahertz radiation experiments, we used as an excitation source an optical parametric amplifier (OPA) pumped by an output of a Ti: sapphire regenerative amplifier (the photon energy of 1.55 eV, the pulse width of 130 fs, and the repetition rate of 1 kHz). The OPA generates a femtosecond laser pulse in the near-infrared region (0.5–1 eV), which is polarized horizontally for 0.5－0.78 eV and vertically for 0.78－1 eV. Figure 2(a) shows the schematic of the experimental setup with the laboratory coordinate (*XYZ*). The plane of incidence is parallel to the [010] direction and the femtosecond laser pulse is introduced to the sample with an incident angle of 45° to the surface normal. The polarization of an incident femtosecond laser pulse is controlled by changing the angle $\theta$ of an quarter waveplate; $\theta$ is defined as an angle of the fast axis in the quarter waveplate relative to the horizontal axis. For 0.5－0.78 eV (0.78－1 eV), the right- and left-handed circularly polarized light are obtained for $\theta$ = 45° or 225° (135° or 315°) and $\theta$ = 135° or 315° (45° or 225°), respectively. On the other hand, for 0.5－0.78 eV (0.78－1 eV), the linearly polarized light along the *X*-direction (*Y*-direction) or equivalently the *p*-polarized (*s*-polarized) light is obtained for $\theta$ = 0°, 90°, 180°, and 270°. The terahertz radiation is

measured in a reflection geometry [Fig. 2(a)]. Unless otherwise stated, we detected the emitted terahertz wave by a standard electro-optic (EO) sampling with an [110]-oriented 1mm-thick ZnTe crystal. By using terahertz radiation imaging [22-26], we confirmed that the BiTeBr crystal used in the present study has a polar single-domain structure. The result of the domain imaging is reported in the Supplemental Material [27]. All the experiments were performed at room temperature.

In the presence of the Rashba effect with a lack of space-inversion symmetry, $\gamma$ in Eq. (1) is expressed as

$$\gamma = \begin{pmatrix} 0 & -\alpha \\ \alpha & 0 \end{pmatrix} \qquad (3).$$

Here, $\alpha$ is a Rashba parameter [2]. By taking into account our experimental setup, in which the plane of incidence includes the [001] direction of the sample, $J_{\text{CPGE}}$ flows along the $Y$-direction [Fig. 2(a)]. This is because $\hat{e}$ directs to the $X$-direction. According to Eq. (2), an irradiation of a circularly polarized femtosecond pulse results in the emission of terahertz waves polarized along the $Y$-direction. Thus, we aimed to detect the $Y$-direction component of the terahertz wave by using a wire-grid polarizer placed in front of the ZnTe detector.

Figure 2(b) shows the electric-field waveforms $E_{\text{THz}}(t)$ of the terahertz radiations observed for the excitations with right- and left-handed circularly polarized light pulses. The photon energy and the fluence of the circularly polarized light pulses are 0.68 eV and 0.17 mJ/cm$^2$, respectively. The time origin $t = 0$ is set at the time when the absolute values of electric-field amplitudes $|E_{\text{THz}}(t)|$ are maxima. A nearly single-cycle electric-field with the temporal width of ~1 ps is observed in each circular polarization, while its phase is inverted to each other [red and blue lines in Fig. 2(b)]. In the upper two panels of Fig. 2(c), we show the corresponding Fourier power spectra, in which the central frequency is 1.0 THz and the spectra expand up to ~2 THz. We also observed the terahertz radiation by the irradiation of a *p*-polarized pulse. The electric-field waveform and its Fourier power spectrum are shown by green lines in Figs. 2(b) and 2(c), respectively, which are almost identical to those of the terahertz radiations induced by the circularly polarized lights, while its central frequency is 1.1 THz and its relative intensity is much smaller.

We next measured the excitation-fluence ($I_{\text{ex}}$) dependence of $E_{\text{THz}}(t)$, when the circularly polarized light is irradiated. Figure 3(a) shows the magnitudes of $E_{\text{THz}}(t)$ at the time origin, $E_{\text{THz}}(0)$, the absolute values of which linearly increase with increasing $I_{\text{ex}}$ up to 0.34 mJ/cm$^2$, and then slightly saturate. Considering these results, we measured all the other data in the condition that $E_{\text{THz}}(0) \propto I_{\text{ex}}$.

Since the sign of the electric-field amplitude is reversed by the switching of the helicity of the incident circularly polarized pulse, the observed terahertz radiation cannot be ascribed to the photo-Dember effect nor the optical rectification, but is attributable to the circular photogalvanic effect. To ascertain this interpretation, we measured the dependence of $E_{\text{THz}}(t)$ on the degree of circular polarization of the incident light pulse (0.68 eV), which can be controlled using quarter-waveplate-angle ($\theta$) [Fig. 2(a)]. Assuming the simple circular photogalvanic effect, $E_{\text{THz}}(0)$ is expected to follow $\sin 2\theta$. Such a $\sin 2\theta$-dependnece is reported in the previous photoconductivity measurements in BiTeBr [28]. The $\theta-$dependence of $E_{\text{THz}}(0)$ is displayed by open circles in Fig. 4. The polarizations of the incident pulse at typical $\theta$ values are illustrated in the upper part in the figure. As can be seen, $E_{\text{THz}}(0)$ reaches the maxima (minima) at $\theta = 50°$ and 230° ($\theta = 120°$ and 305°), which are roughly equal to the angles $\theta = 45°$ and 225° ($\theta = 135°$ and 315°) corresponding to the right-handed (left-handed) circularly polarized lights. Thus, when the helicity of the circularly polarized light is changed the amplitude of the terahertz radiation almost follows $\sin 2\theta$.

More strictly, the observed $\theta$-dependence of $E_{\mathrm{THz}}(0)$ is slightly asymmetric and not completely reproduced by $\sin 2\theta$. This makes us expect the presence of some additional effects. Considering that the terahertz radiations are also caused by the linearly polarized pulse [Fig. 2(b)], a plausible candidate is the linear galvanic effect, in which the effective photocurrent flows along the $Y$-direction without applying the electric field due to the asymmetric scattering of the free carriers in the $k$-space [29,30]. The $\theta$−dependence of $E_{\mathrm{THz}}(0)$ for the linear photogalvanic effect is expected to be $\sin 4\theta$. Using $A \sin 2\theta + B \sin 4\theta$, the observed $\theta$−dependence of $E_{\mathrm{THz}}(0)$ is well reproduced with parameter $A/B = 5.5$ as shown in the purple solid line in Fig. 4. Two components, $A \sin 2\theta$ and $B \sin 4\theta$, are also shown by blue and pink solid lines, respectively. The success of the fitting suggests that the observed terahertz radiation is attributed to the combination of the circular and linear photogalvanic effects.

The circular photogalvanic effect can also be ascertained by changing the photon energy $E_{\mathrm{ex}}$ of the incident pulse. Figure 3(b) shows the $E_{\mathrm{ex}}$−dependence of $E_{\mathrm{THz}}(0)$, which is normalized by the photon number of the incident pulse. $E_{\mathrm{THz}}(0)$ starts to increase at 0.5 eV, which accords with the edge of the interband transition from the Te-5$p$ band to the Bi-6$p$ band [Fig. 1(b)] [20, 30]. Below 0.5 eV, weak absorptions due to the intraband transition exist, which originates from electron carriers introduced in as-grown single crystals due to non-stoichiometry [30]. However $E_{\mathrm{THz}}(0)$ is negligibly small below 0.5 eV, being consistent with the interpretation by the circular photogalvanic effect. This result also rules out the spin-independent photon-drag effect [12, 16], since it usually occurs via the intraband transition. $E_{\mathrm{THz}}(0)$ reaches the maxima at 0.7 eV and finally decreases with increasing the photon energy. The similar tendency was observed in the previous photoconductivity measurements in BiTeBr [28]. For the higher photon energies above 0.7 eV, two kinds of optical transitions can occur in the positive and negative momentum ($\boldsymbol{k}$) spaces with large $|\boldsymbol{k}|$. In this case, photoelectrons with opposite group velocities are simultaneously generated [28]. It results in the cancellation of the photocurrents and the reduction of $E_{\mathrm{THz}}(0)$. Thus, we conclude that the terahertz radiation by the circular polarized pulse in BiTeBr can be attributed to the generation of the spin-polarized photocurrents, which is realized in the asymmetric depopulation of the Dirac state in the subpicosecond time scale.

In narrow-gap semiconductors, the photo-Dember effect is sometimes important as a mechanism of terahertz radiations. However, in our experimental condition in Fig. 2(a) in which the *Y*-polarized component of the terahertz radiation is detected, no terahertz radiation via the photo-Dember effect is detected from the following reason. In general, a surface electric field responsible for the photo-Dember effect is produced along the direction perpendicular to the sample surface. When a light is incident to the sample, photocurrents might be generated along the direction normal to the crystal surface and the resultant terahertz radiation is polarized along the *Z*-direction, which cannot be detected as the *Y*-polarized terahertz radiation but can be detected as the *X*-polarized one in the case of the incident angle of 45° [Fig. 2(a)] (the Supplemental Material [27]).

The optical rectification is another general mechanism for the terahertz radiation in non-centrosymmetric media such as ZnTe [31]. In our results of BiTeBr, a possibility of the terahertz radiation by circularly and linearly polarized lights via the optical rectification can be excluded from the excitation photon-energy dependence. The details are reported in the Supplemental Material [27].

Finally, we discuss the efficiency of the observed terahertz radiation. For this purpose, it is effective to compare the magnitudes of the *Y*-polarized terahertz radiation generated by circularly polarized light with that of the *X*-polarized one generated by the linearly polarized light; the former and the latter originate from the circular photogalvanic effect and from the photo-Dember effect, respectively (the Supplemental Material [27]). The result revealed that the amplitude of the terahertz electric field emitted via the circular photogalvanic effect is approximately 3 times as large as that via the photo-Dember effect. The strong terahertz radiation of the former is attributable to the fact that the asymmetric depopulation of the Dirac state can be induced in a finite range of the *k* space by the irradiation of circularly polarized light, since the up-spin (down-spin) conduction band and the down-spin (up-spin) valence band shift to the opposite *k* directions in BiTeBr as shown in Fig. 1(b).. Thus, our work highlights the importance of the spin-splitting bands in Rashba-type polar semiconductors for the terahertz radiation originating from the spin-polarized photocurrents. Using this phenomenon, the phase of the radiated terahertz waves can be controlled by switching the circular polarization of the incident light without external electric fields. This is a new method for the phase control of terahertz waves, which will be useful in various ways such as sensing, imaging, and spectroscopy.

In summary, we detected the terahertz radiation in a bulk Rashba-type polar semiconductor, BiTeBr, by an irradiation of a femtosecond laser pulse. The phase of the terahertz electric field is reversed by switching the helicity of circular polarization of the incident pulse. The dependence of the amplitude and phase of the terahertz electric field on the polarization, fluence, and photon energy of the incident pulse reveals that the observed radiation is caused by the subpicosecond spin-polarized photocurrents originating from the asymmetric depopulation of the Dirac cone induced by a circularly polarized laser pulse. These results provide a new method to generate the current-induced terahertz radiation, the phase of which can be controlled without external electric fields.

This work was partly supported by a Grant-in-Aid by MEXT (No. 25247058, 25600072, 15K13330, and 16H03847) and CREST (JPMJCR1661 and PMJCR16F2), Japan Science and Technology Agency. Y. K. was supported by Japan Society for the Promotion of Science (JSPS) through Program for Leading Graduate Schools (MERIT) and JSPS Research Fellowships for Young Scientists.

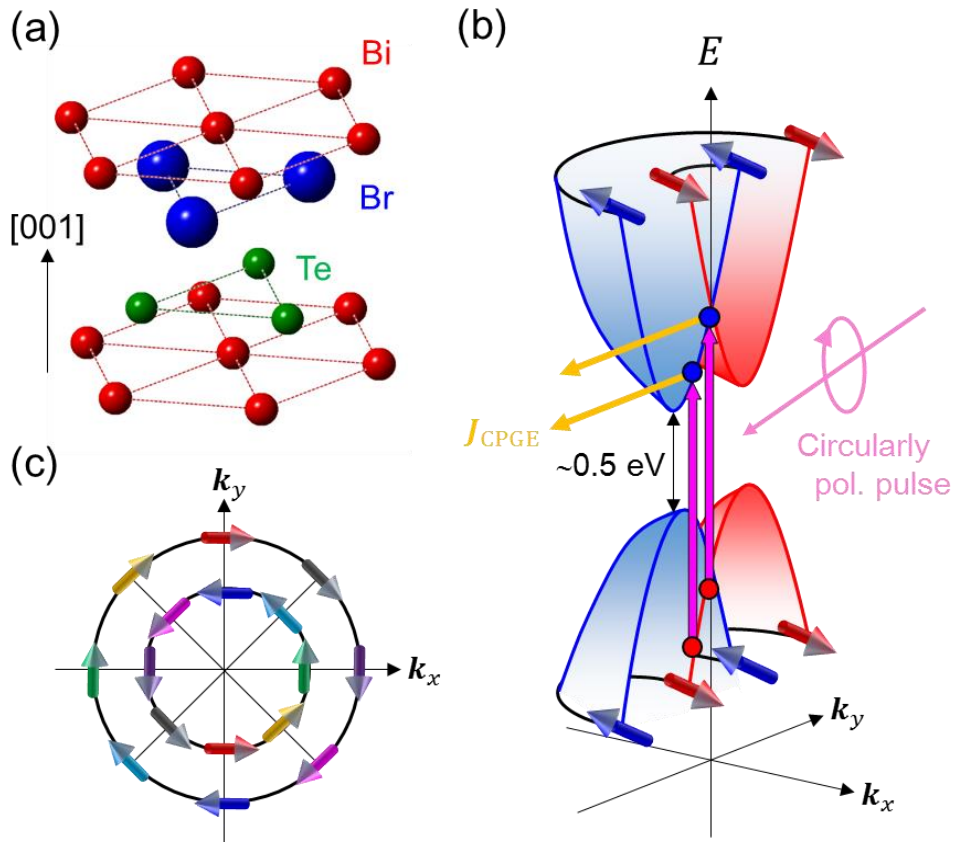

FIG. 1. (a) Crystal structure of BiTeBr with lack of space-inversion symmetry along the [001] direction. (b) Spin-polarized conduction and valance bands by the Rashba-type spin-orbit coupling. (c) Helical spin texture at the Fermi-surface.

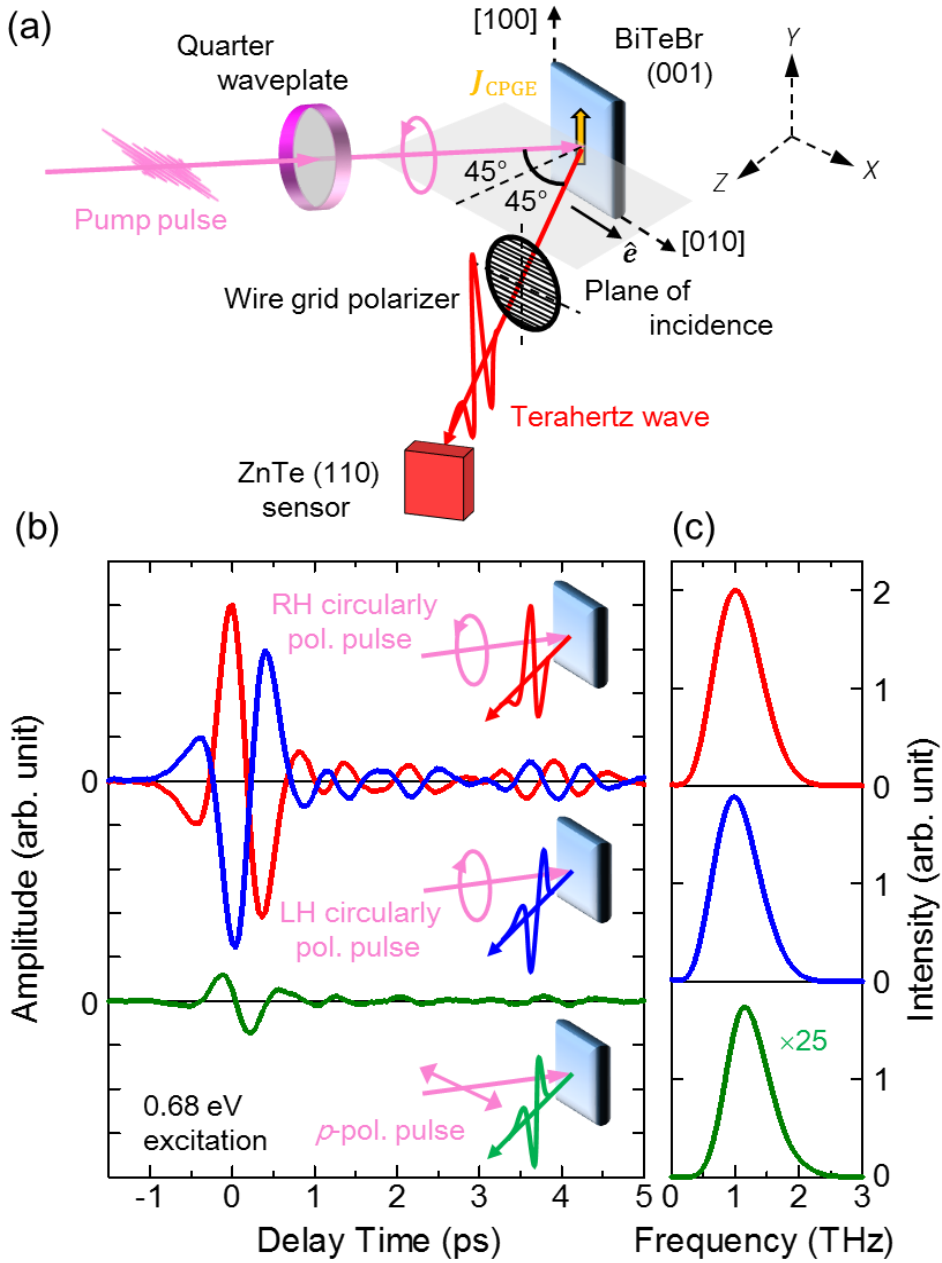

FIG. 2. (a) Schematic of the experimental setup. The polarization state of an incident femtosecond laser pulse was changed by a quarter waveplate. (b) Terahertz waveforms by the irradiation of a femtosecond laser pulse with the right-handed (RH) circular polarization (the red line), the left-handed (LH) circular polarization (the blue line), and the linear ($p$) polarization (the green line). (c) Intensity power spectra obtained by the Fourier transformation of the electric-field waveforms shown in (b).

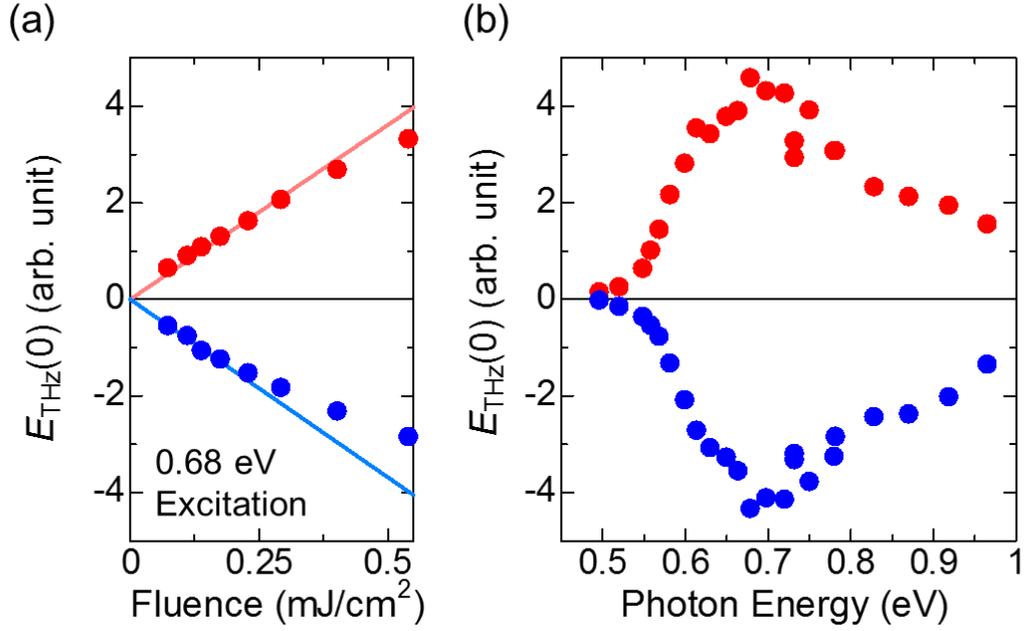

FIG. 3. (a) Excitation fluence ($I_{ex}$) and (b) excitation photon-energy ($E_{ex}$) dependence of the electric-field amplitudes $E_{THz}(0)$ of the terahertz radiations generated by the femtosecond laser pulse with the right-handed circular polarization (red circles) and the left-handed circular polarization (blue circles). The solid lines in (a) indicate the linear relation between $E_{THz}(0)$ and $I_{ex}$.

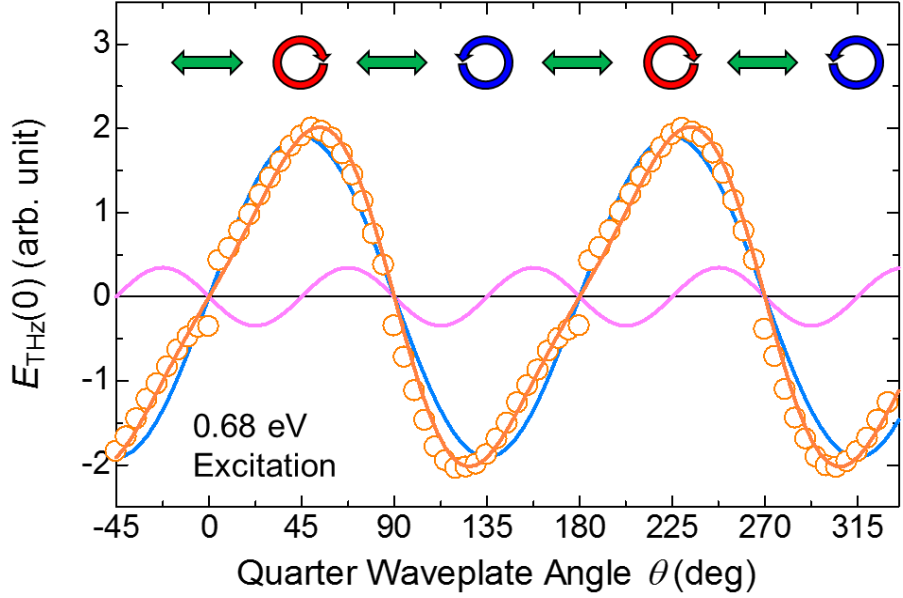

FIG. 4. Quarter-waveplate-angle ($\theta$) dependence of the electric-field amplitudes $E_{\text{THz}}(0)$ of the terahertz radiations at the time origin (open circles). Typical polarization states of the incident femtosecond laser pulse are shown by the upper part in the figure. The pink and orange lines show $\sin 2\theta$ and $\sin 4\theta$ terms in Eq. (3), which originate from the circular and linear photogalvanic effects, respectively (see the text). The fitting of Eq. (3) to the experimental $E_{\text{THz}}(0)$ values is shown by the purple line.

**Supplemental Material**

**Terahertz radiation by subpicosecond spin-polarized photocurrent originating from Dirac electrons in a Rashba-type polar semiconductor**

Yuto Kinoshita[1], Noriaki Kida[1,*], Tatsuya Miyamoto[1], Manabu Kanou[2], Takao Sasagawa[2], and Hiroshi Okamoto[1,3]

[1]*Department of Advanced Materials Science, The University of Tokyo, 5-1-5 Kashiwa-no-ha, Chiba 277-8561, Japan*

[2]*Materials and Structures Laboratory, Tokyo Institute of Technology, Kanagawa 226-8503, Japan*

[3]*AIST-U. Tokyo Advanced Operando-Measurement Technology Open Innovation Laboratory, National Institute of Advanced Industrial Science and Technology, Chiba 277-8568, Japan*

* kida@k.u-tokyo.ac.jp

**Content**

**S1. Domain structure in BiTeBr**

**S2. Terahertz radiation by photo-Dember effect**

**S3. Possibility of terahertz radiation by optical rectification**

## S1. Domain structure in BiTeBr

The previous scanning tunnelling spectroscopy on BiTeI, an analog of BiTeBr, revealed that differently polarized domains coexist on the crystal surface [1]. To map out a structure of polar domains over a sample in ferroelectrics or non-centrosymmetric materials, it is a good way to measure position dependence of electric-field waveforms of terahertz radiations, since the phase of the terahertz electric field depends on the polarization direction. This method was used to visualize not only ferroelectric domains in ferroelectrics [2-5], but also magnetic domains in ferrimagnets [6]. In the present work, by using a femtosecond laser pulse (1771 nm) as the excitation source and a raster scan with a scanning step of 0.4 mm, we measured a spatial image of polar domains over the (001) plane with the size of 1.5 mm × 2.5 mm in a BiTeBr single crystal. We observed no changes in the phase of terahertz electric field on the plane. This ensures the formation of a single-domain structure in BiTeBr.

## S2. Terahertz radiation by photo-Dember effect

In narrow-gap semiconductors, it is well-known that the photo-Dember effect is a dominant terahertz-radiation mechanism. A surface electric field due to the photo-Dember effect is induced along the direction perpendicular to the sample surface. Thus, photocurrents are generated along the same direction when a laser pulse is incident to the sample. This results in the emission of terahertz radiation polarized along the $X$-direction (horizontal direction). In order to compare the magnitude of the $Y$-polarized (vertical direction) terahertz radiation due to the circular photo-galvanic effect [Fig. 2(b)] with that of the $X$-polarized (horizontal direction) one due to the photo-Dember effect, we measured the latter component of the terahertz radiation induced by a $p$-polarized incident light. The measured time-domain waveform and its frequency spectrum are shown in Figs.

S1(a) and (b), respectively. The amplitude of the terahertz electric field emitted via the photo-Dember effect is approximately one third of that emitted via the circular photogalvanic effect [Fig. 2(b)].

**S3. Possibility of terahertz radiation by optical rectification**

BiTeBr is a polar (non-centrosymmetric) semiconductor. The space group of BiTeBr is $P3m1$ and $\chi^{(2)}_{yyy} = -\chi^{(2)}_{yxx} = -\chi^{(2)}_{xyx} = -\chi^{(2)}_{xxy}, \chi^{(2)}_{zxx} = \chi^{(2)}_{zyy} = \chi^{(2)}_{yyz} = \chi^{(2)}_{yzy} = \chi^{(2)}_{xzx} = \chi^{(2)}_{xxz}$, and $\chi^{(2)}_{zzz}$ are nonzero [7]. Therefore, the optical rectification is considered as a possible mechanism of a terahertz radiation [8]. Upon irradiation of a linearly polarized femtosecond laser pulse, the difference frequency generation (DFG) may occur through the modulation of the second-order nonlinear optical polarization $P^{(2)}$. Indeed, this process is dominant in non-centrosymmetric media such as piezoelectric ZnTe and ferroelectric LiNbO$_3$. $P^{(2)}$ is dominated by the second-order nonlinear optical susceptibility tensor $\chi^{(2)}$. An amplitude of a terahertz electric field linearly increases with increasing the coherence length, which is determined by the difference of the speed between the generated terahertz wave and incident light in the medium, and is usually enhanced in the transparent region. However, $E_{\text{THz}}(0)$ in BiTeBr shows the characteristic excitation photon-energy dependence, as shown in Fig. 3(b), and there is no signal in the transparent region below 0.5 eV. Thus, we can consider that the terahertz radiation in BiTeBr is not caused by the optical-rectification process.

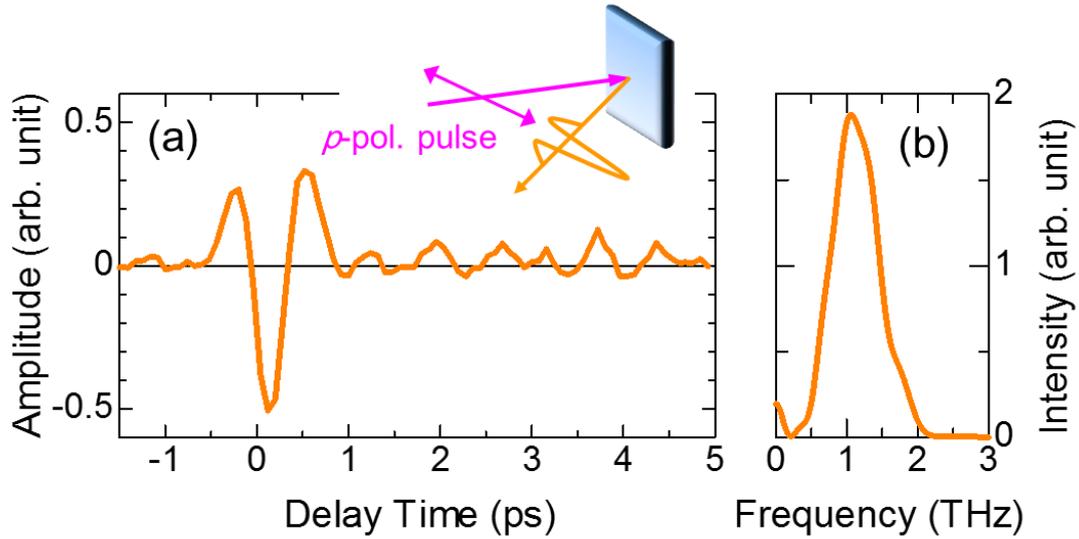

FIG. S1. (a) Terahertz waveform by the irradiation of a femtosecond laser pulse with linear (*p*) polarization, measured in the experimental setup shown in the upper part of the figure. (b) Intensity power spectrum obtained by the Fourier transformation of the electric-field waveforms shown in (a).